\documentclass{webofc}
\usepackage[varg]{txfonts}   
\newcommand{\be}{\begin{equation}}
\newcommand{\ee}{\end{equation}}
\newcommand{\bi}{\begin{itemize}}
\newcommand{\ei}{\end{itemize}}
\newcommand{\ba}{\begin{array}}
\newcommand{\ea}{\end{array}}
\newcommand{\bea}{\begin{eqnarray}}
\newcommand{\eea}{\end{eqnarray}}

\newcommand{\azione}{\mathcal{S}}

\DeclareRobustCommand{\vect}[1]{
  \ifcat#1\relax
    \boldsymbol{#1}
  \else
    \mathbf{#1}
  \fi}
\begin{document}
\title{Holographic Study of the $ Q \bar Q$ Chaotic Dynamics in General Thermal Background}
%
%

\author{\firstname{Nicola} \lastname{Losacco}\inst{1,2}\fnsep\thanks{\email{nicola.losacco@ba.infn.it}}
}

\institute{Istituto Nazionale di Fisica Nucleare, Sezione di Bari,  Via Orabona 4, I-70126 Bari, Italy
\and
           Dipartimento Interateneo di Fisica ``M. Merlin'', Universit\`a  e Politecnico di Bari, \\ via Orabona 4, 70126 Bari, Italy
          }

\abstract{%
  The holographic approach is applied to study the chaotic behaviour of a strongly coupled $Q \bar Q$ pair in general thermal background. We consider two different backgrounds, one with finite temperature and baryon density, and one with finite temperature and constant magnetic field along a fixed direction. The results allow us to understand the dependence of the chaotic dynamics on the background, to test the universal bound on chaos conjectured by Maldacena, Shenker and Standford (MSS).
}
\maketitle
\section{Introduction}
\label{intro}
We study the effects of chemical potential, related to a baryon reservoir, and of a magnetic field on the chaotic behaviour of a strongly coupled $Q \bar Q$ pair in a finite temperature background \cite{Colangelo:2020tpr,Colangelo:2021kmn}. Such systems can be investigated through holographic methods analyzing the dynamics of an open string hanging in the bulk in presence of a black hole (BH). The BH properties, as Hawking temperature and charge, are related to properties of the boundary gauge theory. In \cite{Sekino_2008, susskind2011addendum} it is shown that BH belong to a set of systems called fast ``scramblers'', in particular BH are the fastest scramblers in nature: the time needed for a system near a BH horizon to loose information depends logarithmically on the number of the system degrees of freedom.
As shown in \cite{Maldacena:2015waa} these systems present a bounded chaotic dynamics, with upper bound (MSS bound) on the largest Lyapunov exponent $\lambda$ characterizing the chaotic behaviour of the thermodynamic quantum system with temperature $T$:
	\begin{equation}
\lambda \leqslant 2 \pi T .
\label{eq:1}
\end{equation}
\noindent
The connections between chaotic quantum systems and gravity have been investigated in \cite{Shenker_2014,Shenker:2014cwa,Kitaev,Polchinski:2015cea,Giataganas:2021ghs}.
A generalization of the bound Eq.~\eqref{eq:1}, proposed in the literature for systems presenting a global symmetry \cite{Halder:2019ric}:
\begin{equation}
\lambda \leqslant \frac{2 \pi T}{1-\left|\frac{\mu}{\mu_c}\right|} \qquad \mu \ll \mu_c
\label{eq:2}
\end{equation}
\noindent
where $\mu$ is the chemical potential and $\, \mu_c$ a critical value, can be checked. 
To test the bounds we consider the $Q \bar Q$ pair in a finite temperature and baryon density background \cite{Colangelo:2020tpr}, and with a constant and uniform magnetic field at finite temperature \cite{Colangelo:2021kmn}.
This allows us to test the generalized bound Eq.~\eqref{eq:2}, and the features of phenomenological contexts such as in heavy-ion collisions \cite{Arefeva:2020vae,Arefeva:2022avn,Rannu:2022fxw}. 

\section{Geometry}
\label{sec-1}
The gravity dual of a strongly coupled $Q \bar Q$ pair is an open string hanging in a $5$-dimensional metric obtained solving Einstein equations with suitable boundary conditions. The endpoints of the string are on the boundary, they stand for the heavy quarks in the gauge boundary theory.
For finite baryon density the geometry is described by a Reissner–Nordstrom metric:
\noindent
\begin{equation}
\begin{aligned}
d s^2 &= - f \left( r \right) r^2 d t^2 + r^2 d \vect{x}^2 +\frac{1}{ r^2 f \left( r \right)} d r^2 ,  \\
&f \left( r \right) =  1 - \frac{r_h^4}{r^4} - \frac{\mu^2 r_h^2}{r^4} + \frac{r_h^4 \mu ^2}{r^6},
\label{eq:RNLine}
\end{aligned}
\end{equation}
\noindent
where $r_h$ is the external BH horizon and the chemical potential $\mu$ is related to the BH charge. 
For the case with magnetic field, an approximate solution of the Einstein equations is used \cite{Li:2016gfn}:
\noindent
\begin{align}\label{eq:metricHuang}
    ds^2 = - f \left( r \right) r^2 d t^2 + r^2 h \left( r \right)&( d x^1 )^2 + r^2 h \left( r \right)( d x^2 )^2 + r^2 q \left( r \right) ( d x^3 )^2 + \frac{1}{ r^2 f \left( r \right)} d r^2, \\
&f(r)=1-\frac{r_h^4}{r^4}-\frac{2 B^2}{3 r^4}  \log {\frac{r}{r_h}} \nonumber \\
    &q(r)=1-\frac{2 B^2}{3 r^4}  \log r \\
   &h(r)=1+\frac{B^2}{3 r^4}  \log r\,. \nonumber
\end{align}
\noindent
$B$ is related to the magnetic field in the $x^3$ direction. The magnetic field breaks rotational invariance, hence $h \left( r \right) \neq q \left( r \right)$.

\section{Exploring Chaos}
\label{sec-2}
\begin{figure}[h]
\centering
\includegraphics[width=0.6\linewidth]{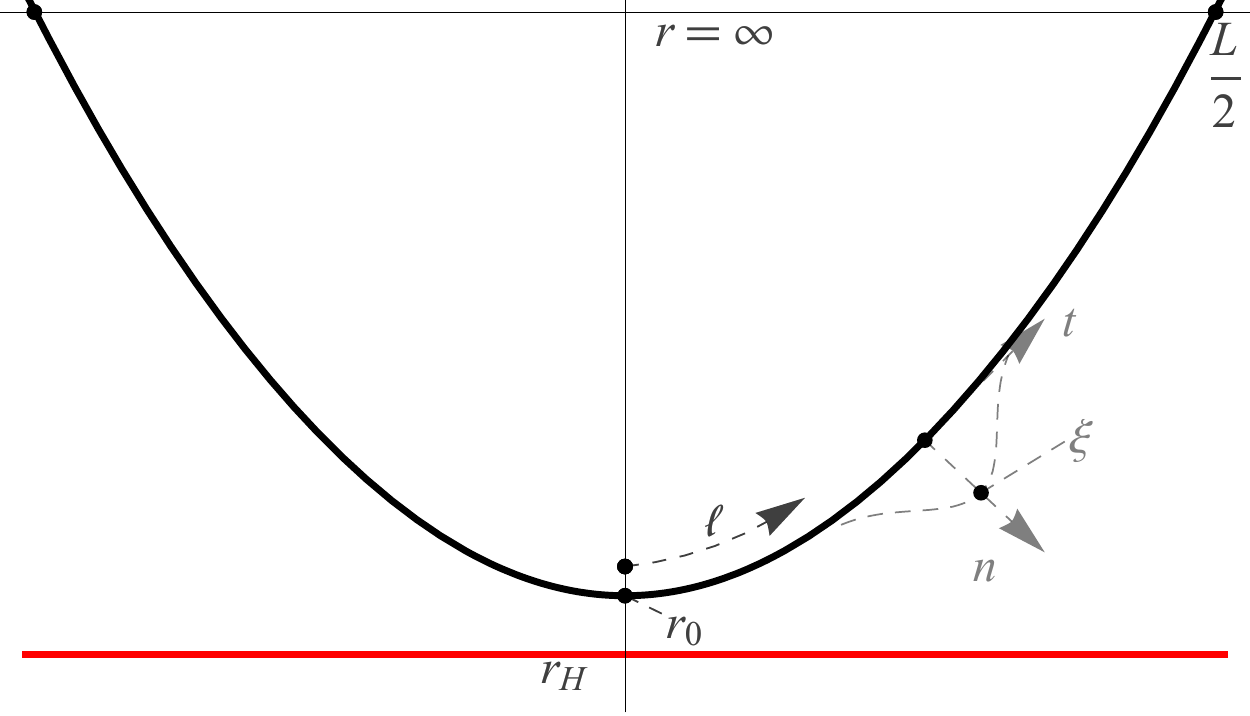}
\caption{Static string profile and perturbation $\xi \left( t, \ell \right)$ along the direction orthogonal to the string in each point with coordinate $\ell$.}
\label{fig-1}       
\end{figure}

Let us consider the general metric
   \noindent
	\begin{equation}
	ds^2=g_{tt} dt^2 + g_{11}(dx^1)^2+ g_{22} (dx^2)^2 + g_{33} (dx^3)^2+g_{rr}dr^2\, .
   \end{equation}
	\noindent
The string profile is obtained from the Nambu-Goto (NG) action:
	\noindent
	\begin{equation}
  	  \azione=-\frac{1}{2\pi \alpha^\prime} 	\int dt\, d \ell \, \sqrt{-h}\,
  	  \label{eq:NG}
	\end{equation}
where $\alpha^\prime$ is the string tension and $h$ the determinant of the induced metric $h_{ij}=g_{MN} \frac{\partial X^M}{\partial \xi_i} \frac{\partial X^N}{\partial \xi_j}$, with $\xi_{i,j}$ the worldsheet coordinates and $g$ the metric tensor.
The static solution depends on the parameters of the metric and on the proximity of the string to the BH horizon, quantified by the string tip position $r_0$. Analyzing the energy of the static configuration dependency on $r_0$, we found that it has a maximum near the black hole horizon $r_h$. This unstable configuration enhances the chaotic behaviour of the system, therefore we set our static solution near the BH horizon.
We perturb the static configuration with a time dependent fluctuation orthogonal in each point of the string \cite{Hashimoto:2018fkb}:
\noindent
\begin{equation}
 \begin{aligned}
r \left( t,\ell \right) &= r_{BG} \left( \ell \right) + \xi \left( t, \ell \right) n^{r} \left( \ell \right),  \\
x\left( t,\ell \right) &= x_{BG} \left( \ell \right) + \xi \left( t,\ell \right) n^{x} \left( \ell \right),
 \end{aligned}
\end{equation}
\noindent
leaving unperturbed the endpoints, Fig.~\ref{fig-1}. $BG$ stands for background solution obtained from the action in Eq.~\eqref{eq:NG}.
\noindent
We expand the action Eq.~\eqref{eq:NG} to the second order in the perturbation $\xi$:
\begin{equation}
\begin{aligned}
S^{\left( 2 \right)} = \frac{1}{2 \pi \alpha^\prime}\int \mathrm{d}t \int_{-\infty}^{\infty} \mathrm{d} \ell \big( C_{tt} \dot{\xi}^2 + C_{\ell \ell}  \acute{\xi}^2 + C_{00} \xi^2\big). 
\end{aligned}
\label{eq:2nd Order Action}
\end{equation}
\noindent
$C_{tt}$, $C_{\ell \ell}$ and $C_{00}$ depend on $\ell$ and on the parameters of the metric. 
From the action Eq.~\eqref{eq:2nd Order Action} we obtain the equation of motion for the perturbation. Factorizing the variables $\xi (t,l) = \xi(l) e^{i \omega t}$, this becames a Sturm-Liouville equation $\partial_\ell \left( C_{\ell \ell} \, \acute{\xi} \right) - C_{00} \, \xi = \omega ^2 C_{tt} \, \xi$.
We solve it for the first two eigenvalues. We write the perturbation as a linear combination of the first two eigenfunctions with time dependent coefficients describing the dynamics of the system:
\begin{equation}
\xi \left( t,\ell \right) = c_0 \left( t \right) e_0 \left( \ell \right) + c_1 \left( t \right) e_1 \left( \ell \right) . 
\label{eq:pert}
\end{equation} 
\noindent 
This allow us to evaluate the third order action using Eq.~\eqref{eq:pert}:
\begin{equation}
 \begin{aligned}
S^{(3)}&=\frac{1}{2 \pi \alpha^\prime}\int \mathrm{d}t \Big[ \sum_{n=0,1}\left(\dot{c}_n^2-\omega_n^2 c_n^2  \right) + K_1 c_0^3 
+ K_2 c_0 c_1^2  + K_3 c_0 \dot{c}_0^2 + K_4 c_0 \dot{c}_1^2 + K_5 \dot{c}_0 c_1 \dot{c}_1\Big].
\label{eq:13}
 \end{aligned}
\end{equation}
\noindent
The coefficients $K_{1, \dots, 5}$ are obtained from an integration over $\ell$ and depend on $r_0$, the tip position of the hanging string, and on the parameters of the metric.
The potential described by Eq. \eqref{eq:13} is a trap potential that confines the dynamics of the unstable string configurations. In the trap region the kinetic term is negative.  As shown in \cite{Hashimoto:2018fkb,Colangelo:2020tpr}, we can substitute $c_{0,1}\to \tilde c_{0,1}$ in the action, with $c_0=\tilde{c}_0 + \alpha_1 \tilde{c}_0^2 + \alpha_2 \tilde{c}_1^2$ and $c_1 = \tilde{c}_1 + \alpha_3 \tilde{c}_0 \tilde{c}_1$, neglecting  $\mathcal{O} ( \tilde{c}_i^4)$ terms, setting the constants $\alpha_i$ ensuring the positivity of the kinetic term. This replacement does not affect the dynamics of the system, and a chaotic behaviour shows up in the transformed system.
The chaotic dynamics can be studied by analyzing the dynamics of $\tilde{c}_0 (t)$ and $\tilde{c}_1 (t)$ governed by the action Eq.~\eqref{eq:13} after the substitution.

\section{Results}
\label{sec-3}

\begin{figure}[h]
\centering		\includegraphics[width=0.5\linewidth]{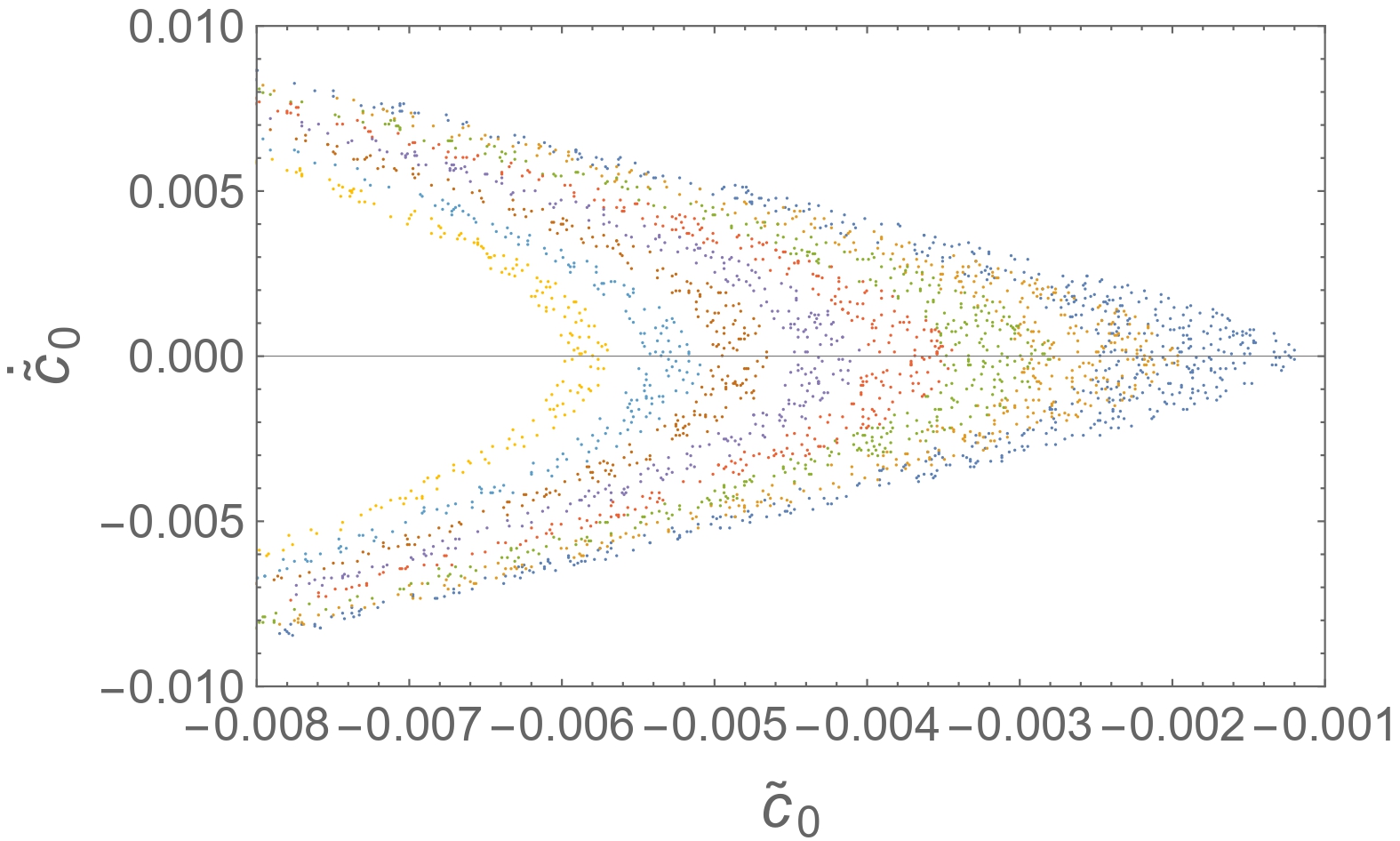}\includegraphics[width=0.5\linewidth]{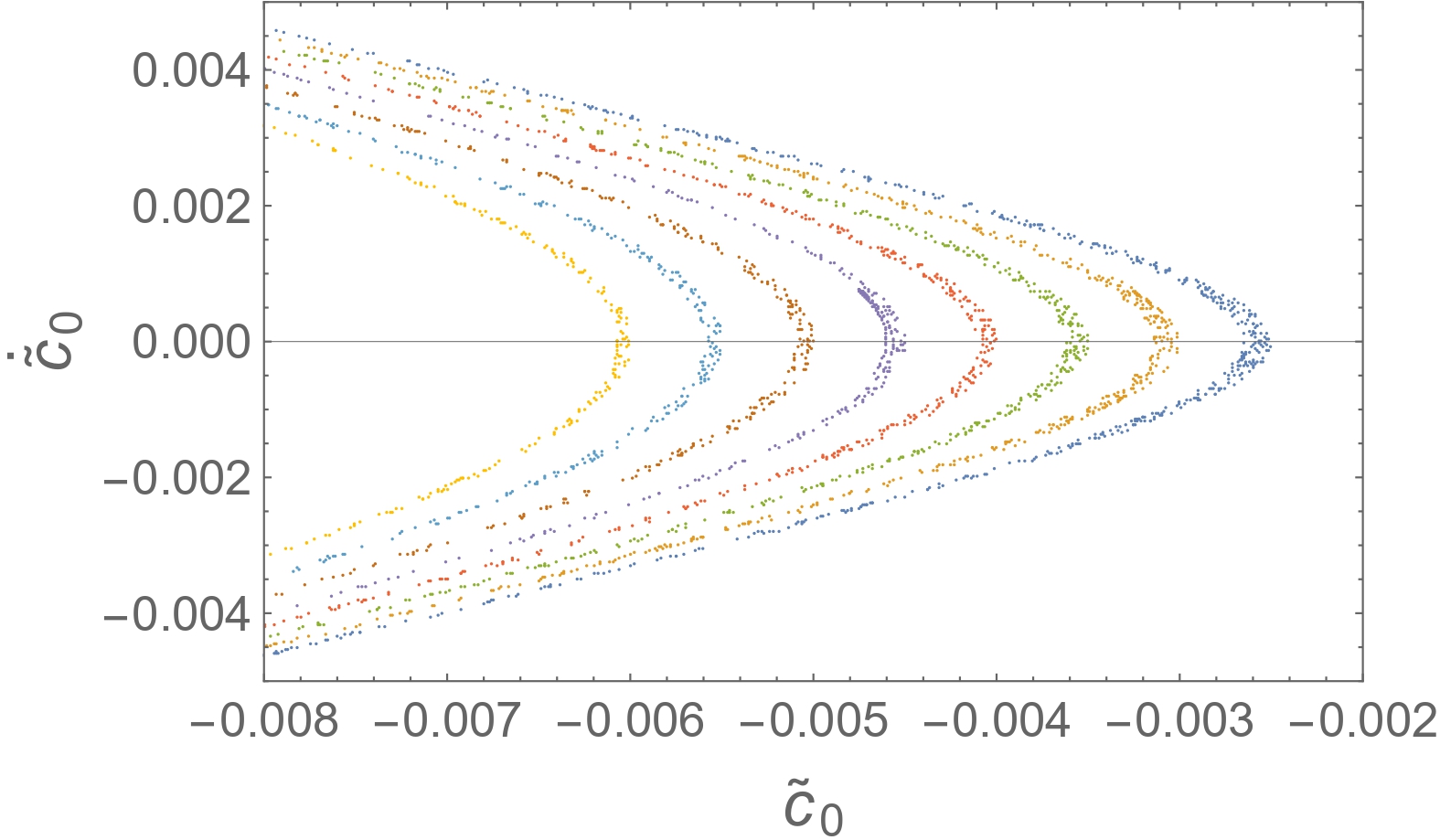}
\caption{Poincar\'e sections for a time-dependent perturbed string, obtained changing the initial conditions, and increasing the chemical potential $\mu =0.3$ (left) and $\mu =0.9$ (right), for $\tilde{c}_1 = 0$ and $\dot{\tilde{c}}_1 \ge 0$.}
\label{fig-2}       
\end{figure}

\begin{figure}[h]
\centering
\includegraphics[width=0.5\linewidth]{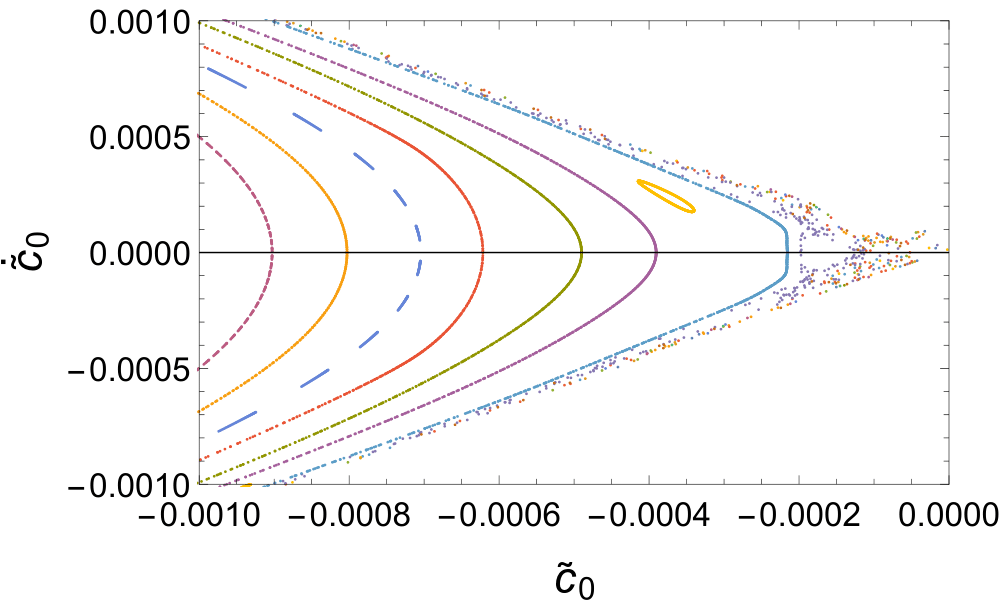}\includegraphics[width=0.5\linewidth]{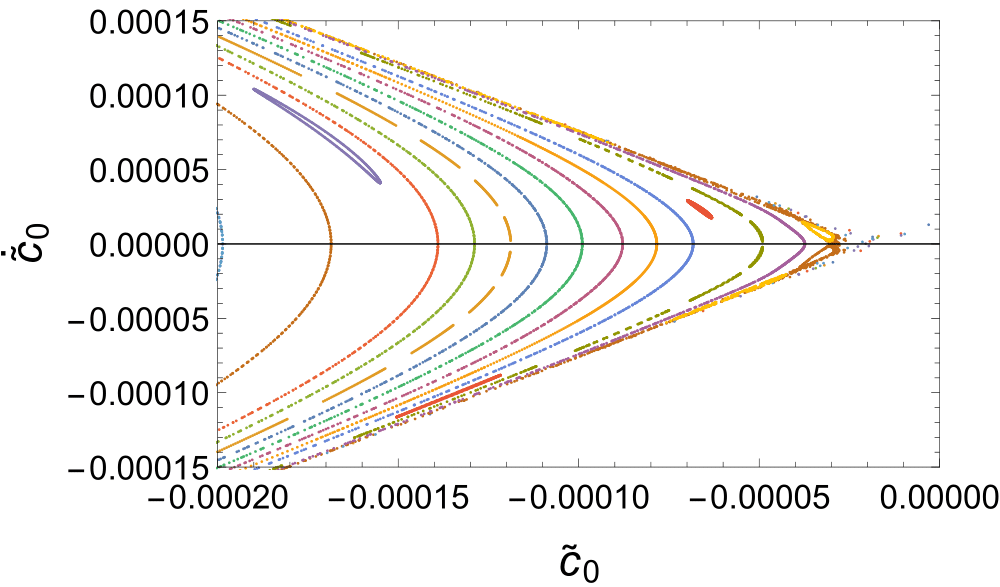} \\
\includegraphics[width=0.5\linewidth]{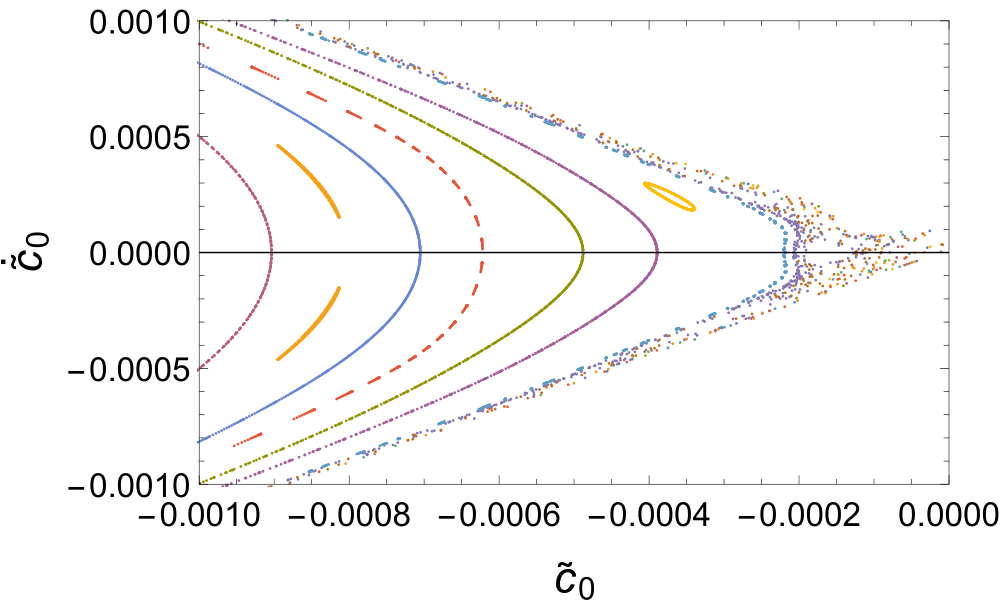}\includegraphics[width=0.5\linewidth]{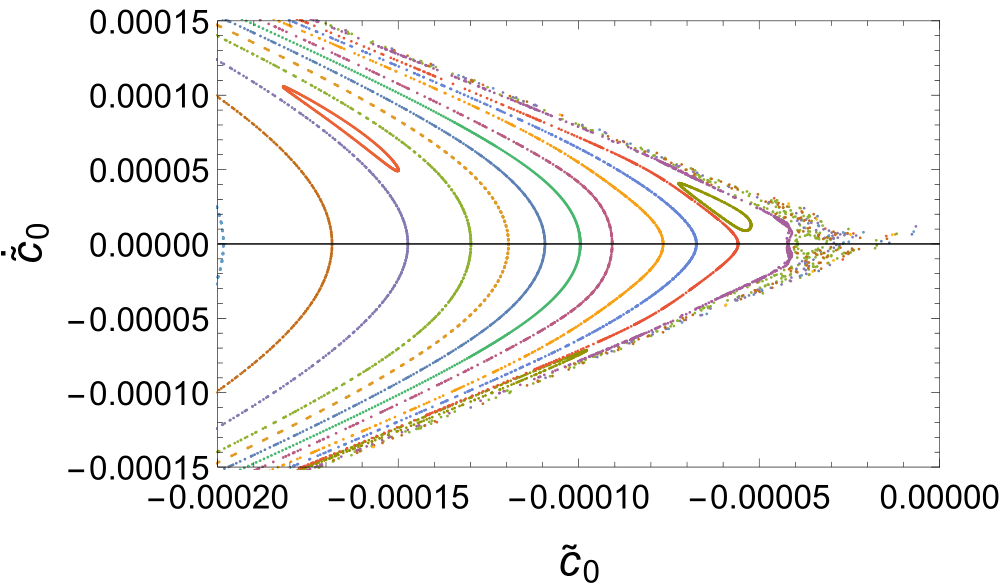}
\caption{Poincar\'e sections for a time-dependent perturbed string along the magnetic field (top panels) and orthogonal to the magnetic field (bottom panels), obtained changing the initial conditions, and increasing the magnetic field $B = 0.3$ (left) and $B = 1$ (right), for $\tilde{c}_1 = 0$ and $\dot{\tilde{c}}_1 \ge 0$.}
\label{fig-3}       
\end{figure}

\begin{figure}[h]
\centering	
\includegraphics[width=0.45\linewidth]{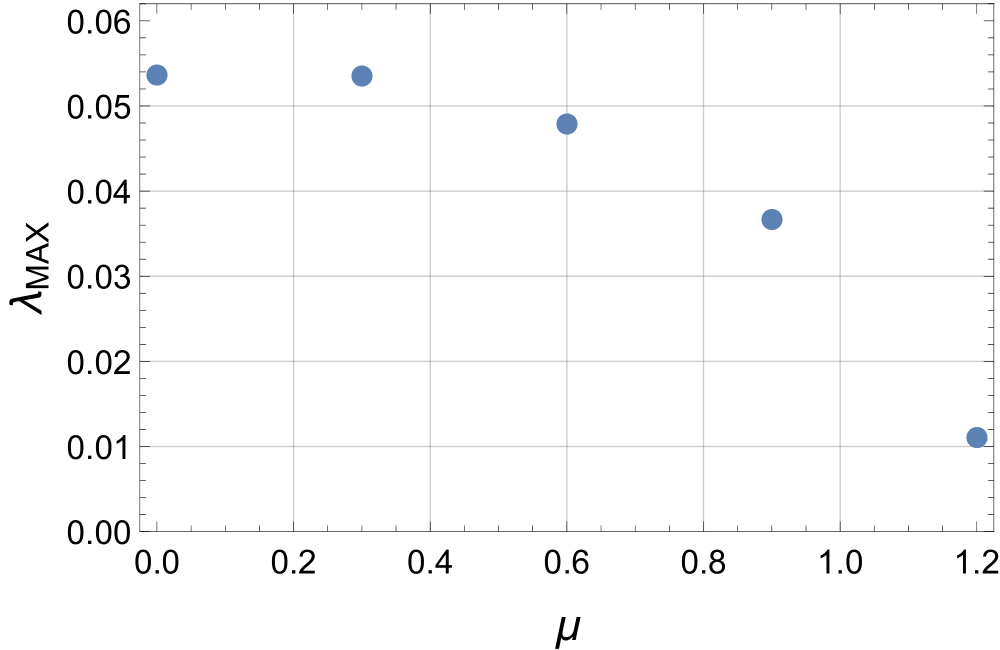}\includegraphics[width=0.49\linewidth]{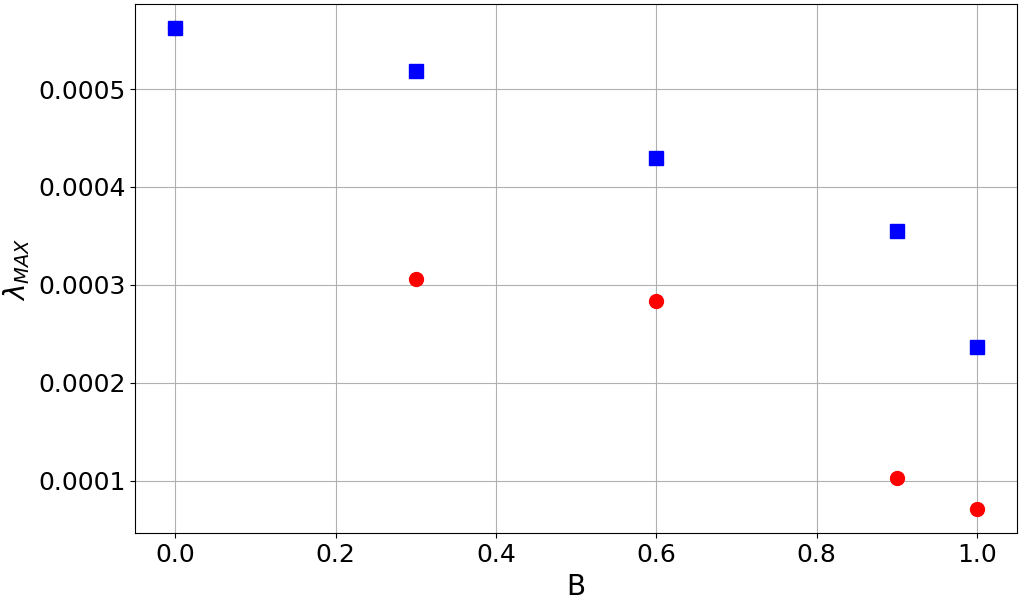}
\caption{Largest Lyapunov exponent  $\lambda_{MAX}$ versus $\mu$ (left) and $B$ (right) for $r_0=1.1$. In the plot at right the results for the string configurations orthogonal (blue squares) and along  (red points) the magnetic field are shown.}
\label{fig-4}       
\end{figure}

The chaotic dynamics is analyzed through Poincar\'e plots and Lyapunov exponents, evaluated at $r_0 = 1.1$ and $r_h=1$ for different $\mu$ and $B$.
In Fig.~\ref{fig-2} we see that at $B=0$ increasing the chemical potential the trajectories in the Poincar\'e section become more stable, hence the chemical potential stabilizes the system.
\noindent
In the case of the magnetic field, either parallel or orthogonal to the $Q \bar Q$ system, we observe a stabilization of the orbits increasing the magnetic field from $B=0.3$ to $B=1$, Fig.~\ref{fig-3}.
\noindent
The anisotropy of the system is manifested in the different chaotic behaviour in the two directions. When the $Q \bar Q$ system is orthogonal to the magnetic field, the region in which the trajectories become scattered is wider than the case in which the string is along the magnetic field, as we can see in Fig.~\ref{fig-3}. It is necessary to approach the origin of the phase space to find chaotic trajectories when the string is along the magnetic field.
All these properties are shown by the largest Lyapunov exponents for different values of chemical potential and magnetic field, Fig.~\ref{fig-4}.
\noindent
Increasing both chemical potential and magnetic field, the systems become less chaotic since the largest Lyapunov exponent decreases. Moreover, the configuration along the magnetic field stabilizes faster. In all cases the largest Lyapunov exponents satisfy the MSS bound Eq.~\eqref{eq:1}, even in the case of baryon density where the generalized bound Eq.~\eqref{eq:2} is less stringent.

\section{Conclusions}
\label{sec-4}

Chaos for the string dual to the $Q \bar Q$ system has been observed in the Poincar\'e plots, characterized by scattered points in the region where the tip of the string is close to the black hole horizon. The system becomes less chaotic increasing  $\mu$ and $B$. For the magnetic field case, an anisotropy effect in two different orientations of the string is found. The MSS bound is satisfied for the largest Lyapunov exponent and remains universal.

\noindent
\\
{\bf Acknowledgements.}
\begin{acknowledgement}
I thank P. Colangelo, F. De Fazio and F. Giannuzzi, co-authors of the works on which this proceeding is based on.
This study has been carried out within the INFN project (Iniziativa Specifica) QFT-HEP.
\end{acknowledgement}
\bibliography{proceeding}
%
%
%
%

\end{document}